\documentclass[pra,aps]{revtex4}

\usepackage{color}

\newcommand{\beq}{\begin{equation}}
\newcommand{\eeq}{\end{equation}}
\newcommand{\bal}{\begin{aligned}[b]}
\newcommand{\eal}{\end{aligned}}
\newcommand{\beqa}{\begin{eqnarray}}
\newcommand{\eeqa}{\end{eqnarray}}

\newcommand{\cblue}{\color{black}}

\begin{document}

\title{Effective Action Approach to Quantum and Thermal Effects: \\
  From One Particle to Bose–Einstein Condensates}

\author{Luca Salasnich$^{1,2,3}$}
\affiliation{$^{1}$Dipartimento di Fisica e Astronomia 'Galileo Galilei' 
  and Padua QTech Center, Universit\`a di Padova,
  via Marzolo 8, 35131 Padova, Italy\\
$^{2}$Istituto Nazionale di Fisica Nucleare, Sezione di Padova, 
via Marzolo 8, 35131 Padova, Italy\\
$^{3}$Istituto Nazionale di Ottica del Consiglio Nazionale delle Ricerche, 
via Carrara 2, 50019 Sesto Fiorentino, Italy}

\begin{abstract}
  We present a detailed derivation of the quantum and
  quantum–thermal effective 
action for non-relativistic systems, starting from the single-particle case 
and extending to the Gross–Pitaevskii (GP) field theory for weakly 
interacting bosons. In the single-particle framework, we introduce the 
one-particle-irreducible (1PI) effective action formalism taking explicitly 
into account the choice of the initial quantum state, its saddle-point 
plus Gaussian-fluctuation approximation, and its finite-temperature 
extension via Matsubara summation, yielding a clear physical interpretation 
in terms of zero-point and thermal contributions to the Helmholtz free 
energy. The formalism is then applied to the GP action, producing the 1PI 
effective potential at zero and finite temperature, including 
beyond-mean-field Lee–Huang–Yang and thermal corrections. We discuss the 
gapless and gapped Bogoliubov spectra, their relevance to equilibrium and 
non-equilibrium regimes, and the role of regularization. Applications 
include the inclusion of an external potential within the local density 
approximation, the derivation of finite-temperature Josephson equations, 
and the extension to D-dimensional systems, with particular attention to the 
zero-dimensional limit. This unified approach provides a transparent 
connection between microscopic quantum fluctuations and effective 
macroscopic equations of motion for Bose–Einstein condensates.
\end{abstract}

\maketitle

\section{Introduction}

Quantum and thermal fluctuations play a crucial role in determining the
properties of many-body systems, particularly in the realm of ultracold atomic
gases and Bose–Einstein condensates (BECs). Since the first theoretical
descriptions of spontaneous symmetry breaking and effective potentials by
Goldstone and Weinberg \cite{goldstone}, and the seminal works by Jona-Lasinio
on effective action methods \cite{jona1,jona2}, the field-theoretic approach
has become a foundational tool to bridge microscopic quantum dynamics and
emergent macroscopic phenomena. Coleman and Weinberg’s pioneering analysis of
radiative corrections in effective potentials \cite{coleman} further enriched
the understanding of fluctuation effects beyond classical approximations
\cite{kleinert,jackiw,abbott}. 
{\cblue For non-relativistic and non-interacting bosons in the presence of
an external potential an effective action approach was developed by Toms and 
Kirsten \cite{toms1,toms2}. For interacting bosons} 
the Gross–Pitaevskii (GP) equation
\cite{gross,pitaevskii} provides a mean-field description
of the condensate wavefunction. However, capturing corrections beyond
mean-field theory, such as the Lee–Huang–Yang quantum fluctuations
\cite{lhy,sala-mgpe,stoof} and thermal
effects \cite{zng,giorgini,sala-cesare}, requires a more
comprehensive framework. The quantum effective action formalism, especially
within the one-particle-irreducible (1PI) scheme \cite{kleinert,jackiw,abbott},
offers a systematic pathway to incorporate these fluctuations by deriving
effective potentials and excitation spectra that reflect both quantum and
thermal contributions. 
Recent experimental advances in ultracold gases have highlighted the
importance of precise theoretical tools for describing weakly interacting
bosons under various conditions, including finite temperature, external
trapping potentials, and low-dimensional regimes
\cite{zng,aybar,pohl}. Moreover, the interplay between
gapless and gapped Bogoliubov modes has implications for equilibrium
properties and non-equilibrium dynamics, such as Josephson
oscillations \cite{josephson,smerzi}, which require refined effective
descriptions \cite{sala-jophase}.

Building on these developments, this work presents a detailed derivation of
the quantum and quantum–thermal effective action for non-relativistic systems.
Starting from the fundamental single-particle case, we clarify the
construction of the 1PI effective action and its approximation
at the Gaussian (one-loop) level. This formalism is then
extended to the GP field theory for weakly interacting bosons, capturing
beyond-mean-field Lee–Huang–Yang corrections and finite-temperature
generalizations. We also discuss the inclusion of external potentials within
local density approximations and the dimensional crossover to analytically
tractable limits, including the zero-dimensional case. 
By providing a transparent and unified framework that connects microscopic
quantum fluctuations to macroscopic condensate dynamics, our approach
complements and extends previous theoretical efforts
\cite{sala-jophase,sala-cesare} and lays the groundwork for
future studies involving dissipative effects, stronger interactions, and
non-equilibrium phenomena.

\section{Single-particle non-relativistic quantum mechanics}

\subsection{Quantum effective action}

Let us consider a particle of mass $m$ and coordinate $q(t)$ described by the
action functional
\beq
S[q(t)] = \int_{t_0}^{t_1} 
\left(  {m\over 2} {\dot q}(t)^2 - V(q(t)) \right) \, dt 
\label{inizioqui}
\eeq
It is well known that the classical trajectory $q_c(t)$ is the one
that extremizes this action, namely
\beq
{\delta S[q_c]\over \delta q(t)} = 0 
\eeq
We now introduce the following decomposition
\beq
q(t) = {\bar q}(t) + \eta(t) 
\label{decomp}
\eeq
where
\beq
{\bar q}(t) = \langle q(t) \rangle
\label{tadpole1}
\eeq
is the average of $q(t)$ with respect some pure or mixed quantum state.
For instance, in the case of a pure quantum state $|\chi\rangle$, such that
$\chi(q,t)=\langle q, t|\chi\rangle$ is its 
wavefunction (normalized to one) in the coordinate representation
at time $t$, we have
\beq
\langle q(t) \rangle = \int_{-\infty}^{+\infty} dq \, q \, |\chi(q,t)|^2 =
\int_{-\infty}^{+\infty} dq \, q \, \big| \int_{-\infty}^{+\infty}
dq_0 \int_{q(t_0)=q_0}^{q(t)=q}
    {\cal D}[q(t')] \, e^{{i\over\hbar}S[q(t')]} \, \chi(q_0,t_0)
    \big|^2
\label{soloio}
\eeq
by using the Feynman path integral representation of the quantum propagator
\beq
K(q,t|q_0,t_0) =\langle q,t|q_0,t_0\rangle = 
\int_{q(t_0)=q_0}^{q(t)=q} {\cal D}[q(t')] \, e^{{i\over\hbar}S[q(t')]} 
\eeq
that gives the conditional probability amplitude of finding
the particle in the position $q$ at time $t$ having it in the position
$q_0$ at time $t_0$, and consequently
\beq
\chi(q,t) = \int_{-\infty}^{+\infty} dq_0 \int_{q(t_0)=q_0}^{q(t)=q}
    {\cal D}[q(t')] \, e^{{i\over\hbar}S[q(t')]} \, \chi(q_0,t_0) 
\eeq
    
From Eq. (\ref{soloio}) one explicitly shows that $\langle q(t)\rangle$
crucially depends on the choice of the wavefunction of the pure quantum
state at the initial time $t_0$. 
Notice that ${\bar q}(t)$ is sometimes called background dynamical variable
and, in general, ${\bar q}(t)\neq q_c(t)$. Thus, usually 
\beq
{\delta S[{\bar q}]\over \delta q(t)} \neq 0 
\eeq
We observe that, as a consequence of Eqs. (\ref{decomp})
and (\ref{tadpole1}), it follows
\beq
\langle \eta(t) \rangle = 0 
\label{tadpole2}
\eeq

The main issue of this paper is to find the functional of ${\bar q}(t)$ whose
Euler-Lagrange equation gives the exact equation of
motion of ${\bar q}(t)$. This functional is usually called
quantum (one-particle irreducible, 1PI) effective action $\Gamma[{\bar q}]$. 
It is important to stress that, in general,
$\Gamma[{\bar q}]$ cannot explicitly
be calculated without doing some approximation or some perturbative expansion.
Usually $\Gamma[{\bar q}]$ is obtained introducing a source term $J(t)$ 
and performing a Legendre transformation \cite{coleman,kleinert}.
Here we will derive the 1PI quantum effective action of the system
without the use of source terms nor Legendre transformations.
In our approach, $\Gamma[{\bar q}]$ is simply given by
\beq
e^{{i\over\hbar}\Gamma[{\bar q}]} = \int {\cal D}[\eta] \, e^{{i\over\hbar}S[{\bar q}+\eta]}
\label{miadefinizione} 
\eeq
provided that Eq. (\ref{tadpole2}) holds. In the absence
of the condition (\ref{tadpole2}), the action of Eq. (\ref{miadefinizione})
is called  background effective action. 

The stationary phase approximation (saddle-point plus Gaussian fluctuations)
of this functional integral around $q={\bar q}$ gives 
\beq
e^{{i\over\hbar}\Gamma[{\bar q}]} \simeq F[{\bar q}] \, e^{{i\over\hbar}
  S[{\bar q}]} 
\eeq
where 
\beq
F[{\bar q}] = {\rm det}\left[ 
  {\delta^2 S[{\bar q}]\over \delta q(t) \delta q(t')} 
  \right]^{-1/2} ={\rm det}\left[
  \left(-m {d^2\over dt^2} - V''({\bar q})\right)
  \, \delta(t-t') \right]^{-1/2} = {\rm det}[{\hat G}]^{-1/2}
\eeq
is the contribution due to Gaussian fluctuations. Quite remarkably,
although expanding $S[{\bar q}+\eta]$ produces a linear term in $\eta$
whenever ${\bar q}(t)$ is not equal to the classical trajectory $q_{c}(t)$,
this contribution does not appear in the effective action $\Gamma[{\bar q}]$.
The reason is that quantum corrections from higher-order fluctuation
terms (tadpoles) exactly compensate the bare linear term, leaving
$\Gamma[{\bar q}]$ free of such contributions by construction.
One can prove \cite{jackiw,abbott} that this compensation is ensured 
by the condition (\ref{tadpole2}), or equivalently by
the definition of the background field (\ref{tadpole1}). 
This guarantees that the linear term in the fluctuation expansion cancels
between the bare action and the quantum tadpole contributions. 

To conclude the discussion, we notice that $F[{\bar q}]$ can be exponentiated:  
\beq
F[{\bar q}] = e^{\ln{(F[{\bar q}])}} 
\eeq
This means that we can write 
\beq
\Gamma[{\bar q}] \simeq S[{\bar q}] +
{\hbar\over i} \ln{(F[{\bar q}])} 
\eeq
Thus, at the Gaussian level, the quantum effective action
$\Gamma[{\bar q}]$ is the classical action $S[{\bar q}]$ plus
the one-loop quantum correction $(\hbar/i)\ln{(F[{\bar q}])}$. 

The exact analytical calculation of $F[{\bar q}]$ can be done only
in the very simple case where
${\bar q}$ in $V({\bar q})$ is time independent. 
If ${\bar q}$ is time dependent, the standard 
trick (lowest-order derivative expansion aka
adiabatic approximation aka local field 
approximation) is to calculate $F[{\bar q}]$ assuming that ${\bar q}(t)$ 
is time independent and restoring the time dependence only at the end
of the calculation. Under this assumption we have 
\beq
\ln(F[{\bar q}]) = - {1\over 2} \ln( {\rm det}[{\hat G}]) = - {1\over 2}
{\rm Tr}[\ln({\hat G})] \simeq - {1\over 2} \int_{-\infty}^{+\infty} dt \,  
\int_{-\infty}^{+\infty} {d\omega\over 2\pi} \,
\ln{\left(\omega^2 - {V''({\bar q}(t))\over m}\right)}  
\label{adiabatic}
\eeq
In this way we identify the effective (one-loop) potential of the system 
\beq
V_{\rm eff}({\bar q}) = V({\bar q}) + {\hbar\over 2i}
\int_{-\infty}^{+\infty} {d\omega\over 2\pi} \,
\ln{\left(\omega^2 - {V''({\bar q})\over m}\right)} 
\eeq
After integration, and discarding spurious divergent terms,  
we obtain 
\beq
V_{\rm eff}({\bar q}) = V({\bar q}) + {\hbar\over 2}
\sqrt{V''({\bar q})\over m}
\label{zero-point}
\eeq
and the corresponding one-loop 1PI action functional
\beq
\Gamma^{\rm (1-loop)}[{\bar q}] = \int 
\left(  {m\over 2} {\dot {\bar q}}^2 - V_{\rm eff}({\bar q})
\right) \, dt 
\eeq

The quantum effective action method is rather complicated but
the final result is quite simple: the quantum effective
action $\Gamma^{\rm (1-loop)}[{\bar q}]$ it is nothing else than
the one derived from the stationary 
phase approximation (saddle point plus Gaussian fluctuations)
of the path integral. The only caveat is that the saddle point  
classical (mean field) solution $q_c(t)$, which extremizes
the classical action $S[q]$, must be re-interpreted as the quantum
average ${\bar q}(t)$ of the dynamical variable $q(t)$ of the problem.
Moreover, the Gaussian beyond-mean-field correction can be interpreted
as the zero-point energy of the harmonic oscillator
of quantum fluctuations with effective frequency
\beq
\omega_{\rm eff}({\bar q}) = \sqrt{V''({\bar q})\over m} 
\label{omegaeff}
\eeq
In conclusion, we observe that it is possible
to extend this effective action
approach to the case of a mass $m$ (see Eq. (\ref{inizioqui})) 
that depends on the dynamical variable $q(t)$: $M(q(t))$. For details
see Ref. \cite{kleinert} and also Refs. \cite{salvatore,sala-cesare+sofia}.

\subsection{Quantum-thermal effective action}

In the spirit of the adiabatic approximation, Eq. (\ref{adiabatic})
can be generalized to the case of finite temperature $T$ as follows
\beqa
\omega &\to& \omega_n = {2\pi \over \hbar \beta} n
\\
\int {d\omega\over 2\pi} &\to& {1\over \hbar\beta} \sum_{n=-\infty}^{+\infty}
\eeqa
where $\omega_n$ are the bosonic Matsubara frequencies and $\beta=1/(k_BT)$
with $k_B$ the Boltzmann constant. Thus, the quantum-thermal effective
potential reads
\beq
V_{\rm eff}^{(T)}({\bar q}) = V({\bar q}) +
{1\over 2i \beta} \sum_{n=-\infty}^{+\infty} 
\ln{\left({4\pi^2n^2\over \hbar^2\beta^2} - {V''({\bar q})\over m}\right)} 
\eeq
namely
\beq
V_{\rm eff}^{(T)}({\bar q}) = V({\bar q}) + {\hbar\over 2}
\sqrt{V''({\bar q})\over m}
+ k_B T \ln\left(1-
\exp{\left(-{\hbar\over k_BT} \sqrt{V''({\bar q})\over m}\right)}\right) 
\label{less-rigorous}
\eeq
This final result has a clear physical interpretation: 
the Gaussian beyond-mean-field quantum-thermal correction is
the Helmholtz free energy of the harmonic oscillator, 
with effective frequency given Eq. (\ref{omegaeff}),  
of quantum-thermal fluctuations around the mean-field (classical) result.
Notice that for $T\to 0$ one has $V_{\rm eff}^{(T)}({\bar q})
\to V_{\rm eff}({\bar q})$. 
The key point is that to obtain an action functional containing 
both time $t$ and temperature $T$ it is useful to have a two times action
functional and then performing a Wick rotation with respect to one of the two
times. Eq. (\ref{adiabatic}) contains implicitly two times: explicitly $t$ and
implicitly $t'$ because $\omega$ is the Fourier dual of $t'$. 

\section{Gross-Pitaevskii quantum field theory}

Let us now face the problem of the non-relativistic quantum field theory
for the Schr\"odinger field $\psi({\bf r},t)$ with action functional
\beq
S[\psi] = \int dt \int d^3{\bf r} \,
\psi^* \left( i\hbar{\partial\over\partial t} +
    {\hbar^2\over 2m}\nabla^2 + \mu - {g\over 2} |\psi|^2 \right) \psi 
\label{action-gpe}
\eeq
where $g$ is the strength of the contact interaction of the 
identical bosonic particles of mass $m$ with chemical potential $\mu$.
The action functional (\ref{action-gpe}) is called Gross-Pitaevskii
(GP) action. Also in this case we are looking for the quantum
effective action of the average ${\bar \psi}({\bf r},t)$ 
of the Schr\"odinger field $\psi({\bf r},t)$. To maintain a connection with the
discussion of the previous section, we denote by
$\mathcal{X}[\psi({\bf r}),t] = 
\langle \psi(\mathbf{r}),t | {\cal X} \rangle$ the normalized wavefunctional
of the field $\psi({\bf r})$ at time $t$, 
where $|\psi(\mathbf{r}),t\rangle$ is the coherent state, i.e. the 
eigenstate, of the quantum field operator ${\hat \psi}({\bf r},t)$
with eigenvalue $\psi({\bf r},t)$,
while $|{\cal X}\rangle$ is a many-body quantum state at the initial time. 
The expectation value
\beq
\bar\psi({\bf r},t) = \langle \psi({\bf r},t)\rangle 
\eeq
of the field $\psi(\mathbf{r},t)$ is computed 
as the average of the configuration field $\psi(\mathbf{r})$ 
with respect to the probability density $|\mathcal{X}[\psi({\bf r}),t]|^2$,
namely   
\beqa
\langle \psi({\bf r},t)\rangle &=& \int \mathcal{D}[\psi({\bf r})]\;
\psi({\bf r}) \;  \big|\mathcal{X}[\psi({\bf r}),t]\big|^2
\nonumber 
\\
&=& \int \mathcal{D}[\psi({\bf r})]\;
\psi({\bf r}) \;
\big| \int {\cal D}[\psi_0({\bf r})]
\int_{\psi({\bf r},t_0)=\psi_0({\bf r})}^{\psi({\bf r},t)=\psi({\bf r})}
    {\cal D}[\psi({\bf r},t')] 
\nonumber
\\
&&
{\cblue 
e^{{i\over \hbar}S[\psi[{\bf r},t']]} \,
{\cal X}[\psi_0({\bf r}),t_0] \big|^2
}
\eeqa
which is a generalization of Eq. (\ref{soloio}).       

Without repeating the procedure developed in the previous section,
taking into account well established results for zero-temperature
Gaussian fluctuations \cite{stoof}, we find that the quantum
effective action is given by 
\beq
\Gamma[\bar{\psi}] = \int dt \int d^3{\bf r} \, \left\{
{\bar \psi}^* \left( i\hbar{\partial\over\partial t} +
{\hbar^2\over 2m}\nabla^2 \right)
{\bar \psi} - V_{\rm eff}({\bar \psi}) \right\} 
\eeq
where
\beq
V_{\rm eff}({\bar \psi}) = - \mu \, |{\bar \psi}|^2 
+ {g\over 2} |{\bar \psi}|^4 + {1\over 2}
\int {d^3{\bf k}\over (2\pi)^3} \, E_k({\bar \psi},\mu) 
\eeq
In this formula appears the zero-point energy of Gaussian quantum
fluctuations, characterized by the gapped Bogoliubov spectrum
\beq
E_k({\bar \psi},\mu) = \sqrt{\left( {\hbar^2k^2\over 2m} - \mu +
  2 g |{\bar \psi}|^2 \right)^2 - g^2 |{\bar \psi}|^2} 
\label{gapped}
\eeq
which reduces to the gapless Bogoliubov spectrum
\beq
E_{k}({\bar \psi}) = \sqrt{{\hbar^2k^2\over 2m}\left({\hbar^2k^2\over 2m} +
  2 g |{\bar \psi}|^2\right)}
\label{gapless}
\eeq
only under the very strong assumption 
\beq
\mu = g |{\bar \psi}|^2 
\eeq
that is justified only at equilibrium. By the way, the Goldstone theorem
(gapless spectrum) works only at equilibrium and at $T=0$. 
The Gaussian beyond-mean-field correction can be interpreted
as the zero-point energy of a set of harmonic oscillators with
effective frequencies $E_k/\hbar$. It is important to stress
that $V_{\rm eff}({\bar \psi})$ is divergent but one can extract
a meaningful finite contribution performing an appropriate regularization,
for details see Ref. \cite{stoof}. In three spatial dimensions 
one gets the so-called Lee-Huang-Yang quantum correction. 

We can easily extend this result at finite temperature. On the basis 
of the physical interpretation of quantum fluctuations as a gas 
of Bogoliubov excitations, by using Eq. (5) of Ref. \cite{sala-cesare} 
we immediately obtain
\beq
V_{\rm eff}^{(T)}({\bar \psi}) =
 - \mu \, |{\bar \psi}|^2 
 + {g\over 2} |{\bar \psi}|^4 +
\int {d^3{\bf k}\over (2\pi)^3} \left[ {1\over 2} E_k({\bar \psi},\mu) 
+ k_B T \ln\left(1-
\exp{\left(-{E_k({\bar \psi},\mu) \over k_B T}\right)}\right) \right] 
\label{effective-gpe}
\eeq
This is the 1PI effective potential of the GP action at finite temperature. 
Indeed, $V_{\rm eff}^{(T)}({\bar \psi})$ is nothing else than the grand
canonical potential $\Omega(\mu,n_0,T)$ of Ref. \cite{sala-cesare} with the
identification $n_0=|{\bar \psi}|^2$. 

The equation of motion of ${\bar \psi}({\bf r},t)$ is given by
\beq
i\hbar{\partial\over\partial t} {\bar \psi} =
- {\hbar^2\over 2m}\nabla^2 {\bar \psi} +
{\partial V_{\rm eff}^{(T)}({\bar \psi})\over \partial {\bar \psi}^*} = 0
\label{tdggpe}
\eeq
where $\mu$ is fixed by imposing that
\beq
N_0 = \int d^3{\bf r} \, |{\bar \psi}({\bf r},t)|^2 
\eeq
Our Eq. (\ref{tdggpe}) reduces to the stationary generalized Gross-Pitaevskii
equation discussed in Refs. \cite{aybar,pohl} setting 
${\partial {\bar \psi}\over \partial t}=0$ and adopting the
spectrum (\ref{gapped}) instead of (\ref{gapless})
in $V_{\rm eff}^{(T)}({\bar \psi})$.

{\cblue The standard approach is to adopt the gapless Bogoliubov
 spectrum in Eq. (\ref{tdggpe}) under the assumption of working near thermal
equilibrium}. Introducing the local number density
\beq 
n_0({\bf r},t)=|{\bar \psi}({\bf r},t)|^2 
\eeq
Eq. (\ref{tdggpe}) can be then written as
\beq
i\hbar{\partial\over\partial t} {\bar \psi} =
\left[ - {\hbar^2\over 2m}\nabla^2 -\mu
  + g \, n_0 + \mu_{\rm LHY}(n_0) + \mu_{\rm th}(n_0,T) \right]{\bar \psi}
\label{tdggpe0}
\eeq
where
\beq
\mu_{\rm LHY}(n_0) = g \, n_0 {32\over 3\sqrt{\pi}} \, (n_0 a_s^3)^{1/2} 
\eeq
is the renormalized zero-temperature Lee-Huang-Yang beyond-mean-field 
correction,
i.e. a sort of additional bulk chemical potential with $a_s$ the s-wave
scattering length such that $g=4\pi\hbar^2a_s/m$, while 
\beq
\mu_{\rm th}(n_0,T) = 2 \, g \, n_{\rm th}(n_0,T)
\eeq
is the quantum-thermal correction, where 
\beq
n_{\rm th}(n_0,T) = \int {d^3{\bf k}\over (2\pi)^3}
\left({\hbar^2k^2\over 2m} + g n_0\right){1\over E_k(n_0)}
     {1\over e^{E_k(n_0)/(k_BT)}-1}
\label{nth}
\eeq
plays the role of the density of a thermal bath, such that
\beq
n_{\rm th}(n_0,T) \to \zeta(3/2) \left({m k_B T\over 2\pi \hbar^2}\right)^{3/2}
\quad\quad \mbox{for $n_0$ very small}
\eeq
with $\zeta(3/2)\simeq 2.612$ and
\beq
n_{\rm th}(n_0,T) \to {\zeta(3)\over\pi^2}
\left({k_B T\over \hbar}\right)^3 \left({m\over g n_0} \right)^{3/2} 
\quad\quad \mbox{for $n_0$ very large}
\eeq
with $\zeta(3)\simeq 1.202$. We stress that Eq. (\ref{tdggpe0})
can also be seen as the time-dependent extension of the stationary
Gross-Pitaevskii equation for the Bose-Einstein condensate 
which appears in the Zaremba-Nikuni-Griffin (ZNG) 
formalism \cite{zng}. By adopting the ZNG formalism we recover
our thermal density $n_{\rm th}$ from a Boltzmann equation 
of non-condensed bosons. At equilibrium and zero temperature 
the Lee-Huang-Yang term of Eq. (\ref{tdggpe0}) is consistent
with the modified Gross–Pitaevskii equation of Ref. \cite{sala-mgpe}.

\subsection{Including an external potential}

In the spirit of the local density approximation (LDA) we can also consider 
the inclusion of an external potential $U({\bf r})$
in Eq. (\ref{tdggpe0}), namely
\beq
i\hbar{\partial\over\partial t} {\bar \psi} =
\left[ - {\hbar^2\over 2m}\nabla^2 + U({\bf r}) -\mu
  + g \, n_0 + \mu_{\rm LHY}(n_0) + \mu_{\rm th}(n_0,T) \right]{\bar \psi} 
\label{tdggpe1}
\eeq
Remember that here, and in Eq. (\ref{tdggpe0}),
the chemical potential $\mu$ can be formally removed with an
appropriate redefinition of the phase of time-dependent 
field ${\bar \psi}({\bf r},t)$.

A drawback of Eq. (\ref{tdggpe1}) is that,
for $g\to 0$ the thermal density $n_{\rm th}$ becomes uniform despite
the presence of $U({\bf r})$.
To cure this drawback it is sufficient to make the substitution
$g n_0\to g n_0 + U({\bf r})$ in Eq. (\ref{nth}). However, just to simplify
a bit the problem one usually considers the following Hartree approximation
of the Bogoliubov spectrum in the single-particle phase space \cite{giorgini}
\beq
E_k^{\rm (HF)}({\bf r},t) = {\hbar^2k^2\over 2m} + U({\bf r}) + g \, n_0({\bf r},t)
\eeq
that is reliable for large $k$. In this way we have
\beq
i\hbar{\partial\over\partial t} {\bar \psi} =
\left[ - {\hbar^2\over 2m}\nabla^2 + U({\bf r}) -\mu
  + g \, n_0 + \mu_{\rm LHY}(n_0) + 2\, g\,
  {\tilde n}_{\rm th}(n_0,T) \right]{\bar \psi}
\label{pera1}
\eeq
with
\beq
{\tilde n}_{\rm th}(n_0,T) = \int {d^3{\bf k}\over (2\pi)^3}
{1\over e^{( {\hbar^2k^2\over 2m} + U({\bf r}) + g \, n_0({\bf r},t))/(k_BT)}-1} 
\label{pera2}
\eeq
At this point an important remark is needed. Depending on the
adopted formalism slightly different versions
of Eqs. (\ref{pera1}) and (\ref{pera2}) are derived.
For instance, in the ZNG approach $n_0$ is substituted by
the total number density $n$ in the LHY chemical potential
and also in the thermal density. 

\subsection{Josephson equations at finite temperature}

Under the assumption of an external potential $U({\bf r})$
which separates our three-dimensional ($D=3$) system
in two weakly-linked regions 
and that ${\bar \psi}({\bf r},t)$ of Eq. (\ref{tdggpe1}) 
truly describes Bose-condensed particles at finite temperature $T$, 
a straightforward generalization of the Josephson equations 
\cite{josephson,smerzi} at finite temperature reads
\beqa
\hbar \, {\dot z} &=& - 2 \, J \sqrt{1-z^2} \sin(\phi) 
\\
\hbar \, {\dot \phi} &=& 2 \left( f_T\big[{\bar{n}_0\over 2}(1+z)\big]
- f_T\big[{\bar{n}_0\over 2}(1-z)\big]
\right) + 2\, J {z\over \sqrt{1-z^2}} \cos(\phi) 
\eeqa
where $z(t)$ is the population imbalance of the Bose-Einstein condensate,
$\phi(t)$ is the relative phase of the Bose condensate,
$J$ is the tunneling energy of condensed bosons, ${\bar n}_0$ is the
space-time independent average number density of condensed bosons in the
two regions, and 
\beq
f_T[x] =  g \, x + \mu_{\rm LHY}(x) + \mu_{\rm th}(x,T) 
\eeq
Working with a fixed total average number density ${\bar n}$
of bosons, the average number density ${\bar n}_0$ of
condensed bosons can be extracted from the {\cblue finite-temperature}
Bogoliubov formula
\beq
    {\bar n}_0 = {\bar n} - {8\over 3\sqrt{\pi}} (a_s {\bar n})^{3/2}
    - \bar{n} \, I[a_s {\bar n}^{1/3},{mk_BT\over \hbar^2 \bar{n}^{2/3}}]
    \label{n01}
\eeq
{\cblue with $I[x,y]$ given by
\beq
I[x,y] = {1\over 24 \sqrt{\pi}}{y^2\over x^{1/2}}
\label{n02}
\eeq
A detailed derivation of Eq. (\ref{n01}) with Eq. (\ref{n02}) is
discussed in Ref. \cite{sala-cesare}. The main idea is to write
the thermodynamic grand potential $\Omega$ of the system at equlibrium
as a function of both the chemical potential $\mu$ and the condensate
density ${\bar n}_0=|\bar{\psi}|^2$ by using the gapped
Bogoliubov spectrum (\ref{gapped}). The number density is then obtained as
${\bar n}=-{\partial \Omega\over\partial \mu}{1\over L^3}$. Finally, setting
$\mu = g {\bar n}_0$ one finds ${\bar n}$ as a function
of ${\bar n}_0$ and $T$.}  

\subsection{1PI GP effective potential in D dimensions}

The generalization of Eq. (\ref{effective-gpe}) to the case
of $D$-dimensional bosonic system is immediate:  
\beq
V_{\rm eff}^{(T)}({\bar \psi}) =
 - \mu \, |{\bar \psi}|^2 
 + {g\over 2} |{\bar \psi}|^4 +
\int {d^D{\bf k}\over (2\pi)^D} \left[ {1\over 2} E_k({\bar \psi},\mu) 
+ k_B T \ln\left(1-
\exp{\left(-{E_k({\bar \psi},\mu) \over k_B T}\right)}\right) \right] 
\eeq
As previously discussed, for $D\neq 0$, we can safely use
the gapless spectrum $E_k({\bar \psi},\mu=g|{\bar \psi}|^2)$ instead of
the gapped spectrum $E_k({\bar \psi},\mu)$
in this effective potential. However, this substitution cannot be done
for $D=0$, where only the mode ${\bf k}={\bf 0}$ survives, because
it will imply $E_k=E_0=0$, see Eq. (\ref{gapless}).

In the zero-dimensional case ($D=0$) one must use the gapped spectrum
of Eq. (\ref{gapped}) with $k=0$, namely
\beq
E_0({\bar \psi},\mu) = \sqrt{\left(\mu - 
  2 g |{\bar \psi}|^2 \right)^2 - g^2 |{\bar \psi}|^2} 
\eeq
Thus, the 1PI GP effective potential for $D=0$ reads
\beq
V_{\rm eff}^{(T)}({\bar \psi}) =
 - \mu \, |{\bar \psi}|^2 
 + {g\over 2} |{\bar \psi}|^4 +
{1\over 2} E_0({\bar \psi},\mu) 
+ k_B T \ln\left(1-
\exp{\left(-{E_0({\bar \psi},\mu) \over k_B T}\right)}\right) 
\eeq

Actually, a better treatment at $T=0$ is obtained
considering the gapless Bogoliubov spectrum with a finite ${\bf k}$ and
dimensional regularization around $D=0$ \cite{bardin}.
At finite temperature one instead uses the gapped $E_0({\bar \psi},\mu)$.
At the end, we get
\beq
V_{\rm eff}^{(T)}({\bar \psi}) =
 - \mu \, |{\bar \psi}|^2 
 + {g\over 2} |{\bar \psi}|^4 - {g\over 2} |{\bar \psi}|^2 
+ k_B T \ln\left(1-
\exp{\left(-{E_0({\bar \psi},\mu) \over k_B T}\right)}\right) 
\eeq
Quite remarkably, the term $(g/2)(|{\bar \psi}|^4 - |{\bar \psi}|^2)=
(g/2)|{\bar \psi}|^2 ( |{\bar \psi}|^2-1)$ gives for $T=0$ the exact 
internal energy $(g/2)N(N-1)$ at equilibrium with $N=|{\bar \psi}|^2$.

\section{Conclusions}

We have reviewed and extended the quantum effective action formalism for
non-relativistic systems, with emphasis on its Gaussian (one-loop)
implementation and finite-temperature generalization. Starting from the
single-particle case, we showed how the 1PI effective action can be 
systematically derived and how the saddle-point plus Gaussian-fluctuation
approximation leads to simple yet physically transparent
expressions for the effective potential. The finite-temperature extension,
obtained through Matsubara summation, yields quantum–thermal
corrections with a clear interpretation as the Helmholtz free energy
of harmonic modes with effective Bogoliubov frequencies. 
Applied to the Gross–Pitaevskii field theory, this approach naturally
incorporates beyond-mean-field effects, including the Lee–Huang–Yang
correction and its finite-temperature counterpart, and accommodates both
gapped and gapless excitation spectra. The formalism is flexible enough to
handle external potentials within the local density or Hartree approximations,
to describe Josephson dynamics at finite temperature, and to generalize to
arbitrary spatial dimension, including the analytically interesting
zero-dimensional limit. 
Overall, the quantum–thermal effective action framework offers a unified and
physically intuitive route from microscopic fluctuations to macroscopic
dynamical equations for Bose–Einstein condensates. It connects
field-theoretic principles with experimentally relevant phenomena, providing
a solid base for further developments, such as the inclusion of dissipative
effects, stronger interaction regimes, or non-equilibrium dynamics.
{\cblue Regarding the connection with experiments, it is important to stress
  that, in the context of superconducting Josephson junctions,
  the zero-point correction of Eq. (\ref{zero-point}) explains the shift of the
  quantized energy levels of qubits \cite{devoret}. Instead, in the context
  of ultracold bosonic atoms, the beyond-mean-field Gaussian quantum corrections
  described here have been found in serveral experiments (see, for instance,
  Refs. \cite{papp,wild,cabrera}). In this review we have not
  considered open quantum systems and the impact of external environments.
  This is an hot topic of research where, however, a clear connection between 
  the formalism of master equations \cite{master1,master2}
  and the quantum effective action approach is still lacking.}

\noindent
    {\bf Acknowledgements}:
    The author thanks Francesco Ancilotto, Alessandro Pennacchio,
Sofia Salvatore, and Cesare Vianello for useful discussions and suggestions.

\noindent
{\bf Author Contributions}: This is a single-author paper. 

\noindent
    {\bf Funding}: PRIN 2022 Project of MUR "Quantum Atomic Mixtures: Droplets,
Topological Structures, and Vortices" (2023-2024); Dipartimenti di Eccellenza
Project of MUR ``Frontiere Quantistiche''; Iniziativa Specifica ``Quantum''
of INFN.

\noindent
{\bf Data Availability Statement}: Not applicable. 

\noindent
    {\bf Conflicts of Interest}:
    The authors declare no conflicts of interest.

\end{document}